\documentclass[aps,prl,twocolumn,groupedaddress,showpacs,showkeys]{revtex4}
\usepackage{graphicx}
\begin{document}

\title{Fabrication of closely spaced, independently contacted  Electron-Hole bilayers
in GaAs-AlGaAs heterostructures}

\author{J. A. Keogh}\author{K. Das Gupta}\email{kd241@cam.ac.uk}\author{H. E. Beere}\author{D. A. Ritchie}\author{M. Pepper}
\affiliation{Cavendish Laboratory, University of Cambridge, Madingley Road, Cambridge CB3 0HE, UK.}

\begin{abstract}
We describe a technique to fabricate closely spaced electron-hole bilayers in GaAs-AlGaAs
heterostructures.
Our technique incorporates a novel method for making shallow contacts to a low density
($<10^{11}cm^{-2}$) 2-dimensional electron gas (2DEG) that do not require annealing.
Four terminal measurements on both layers (25nm apart)
are possible.
Measurements show a hole mobility $\mu_{h}>10^{5}{\rm cm}^{2}{\rm V}^{-1}{\rm s}^{-1}$  and an electron
mobility $\mu_{e}>10^{6}{\rm cm}^{2}{\rm V}^{-1}{\rm s}^{-1}$ at 1.5K.
Preliminary drag measurements made down to T=300mK
indicate an  enhancement of coulomb interaction over the values obtained from a
static Random Phase Approximation (RPA) calculation.

\end{abstract}
\date{\today}

\maketitle

Advances in MBE growth techniques over the last two decades have made possible fabrication of
closely spaced double quantum well structures. Such 2$\times$2DEG (double 2D electron gas)
and 2$\times$2DHG (double 2D hole gas) structures
have enabled the most definitive measurements of quantities like
electron-electron scattering rates, compressibility
etc.,\cite{gramilla,eisenstein1,millard} where many-body effects play a crucial role.
Several interesting possibilities\cite{ vignale,balatsky,thakur} have been discussed about the
formation of correlated phases with superfluid-like properties, in electron-hole bilayers,
when the interlayer spacing becomes
comparable to the electron-hole Bohr radius ($\sim$15nm in GaAs). The $\nu$=1 bilayer state in
2$\times$2DEG and 2$\times$2DHG emulates a true electron-hole bilayer in certain ways. Recent
experiments\cite{eisenstein2,kellog,tutuc} on these have shown a remarkable enhancement of the
Hall drag.
However compared to these structures, an electron-hole bilayer is much more difficult to fabricate
and contact.
If modulation doping is used to populate both quantum wells, a bend bending
of 1.5V (bandgap of GaAs) over $\sim$15nm must exist in the barrier layer. The built-in
electric field of $\sim$10$^{8}$V/m, must arise self-consistently from the ionised dopants and the
free carriers. Devices in which simultaneous accumulation of
electrons and holes in close proximity has been demonstrated, \cite{sivan,kane,pohlt,shapira} reduce the
amount of band-bending required by
introducing a discontinuity in the electrochemical potential across
the barrier.
Considering the novel states which are thought to exist in electron-hole bilayers, a robust fabrication
process that lends itself fully to conventional MBE growth and lithographic processing, would be of
considerable importance.

\begin{figure}[h]
\begin{center}
\includegraphics[width=8.5cm,clip]{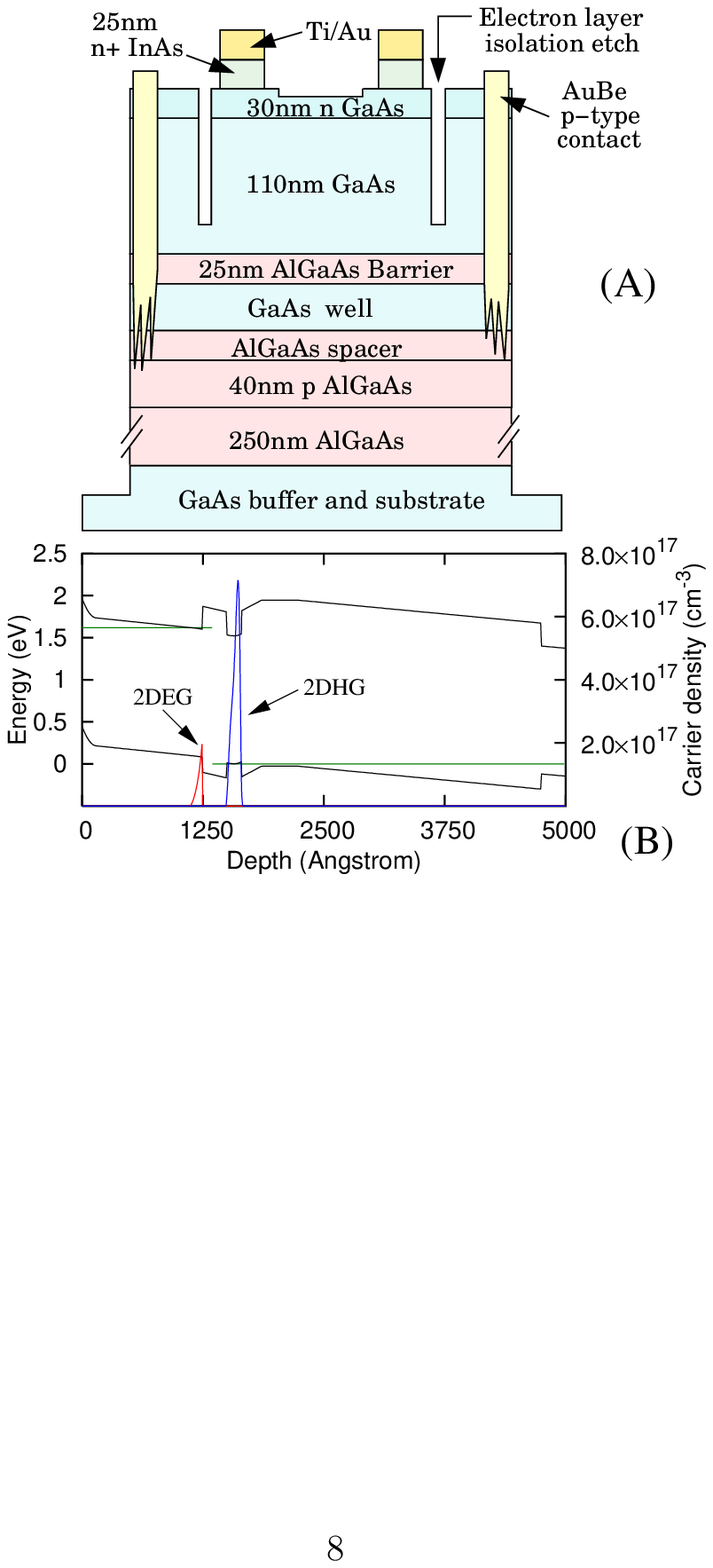}
\end{center}
\caption{\label{deviceschematic}(Color online)
(A) Schematic of the device and (B) self-consistent band structure of a typical device, away from the
InAs contacts. An interlayer bias of 1.62V has been assumed.}
\end{figure}

In this letter, we describe an electron-hole device (see Fig\ref{deviceschematic}) with a
25nm Al$_{0.3}$Ga$_{0.7}$As barrier. Both quantum wells (QW) are in the $\lq\lq$inverted" configuration. The lower hole QW is doped. Electrons
accumulate in the upper QW  under biasing. We use a novel scheme of shallow contacts using a
heavily doped (8$\times10^{18}$cm$^{-3}$ Si) GaAs/InAs capping
layer to contact the electron layer. Doped InAs
pins the Fermi level {\it above} the conduction band at the surface. Any metal which adheres
well to the surface (e.g. Ti/Au) forms an ohmic contact to the region below.
A nearly flatband condition is maintained in the region between the contact and
the 2DEG, allowing a  $\lq\lq$non-spiking" ohmic contact to the 2DEG.
These contacts do not require annealing and have been found to work down to 2DEG
densities in the low 10$^{10}$cm$^{-2}$ range till T$<$300mK. Unlike ion-implanted contacts, a
high temperature anneal ($\sim$800C) is not required to activate them.

\begin{figure}[h]
\vspace*{\fill}
\begin{center}
\includegraphics[width=7cm,clip]{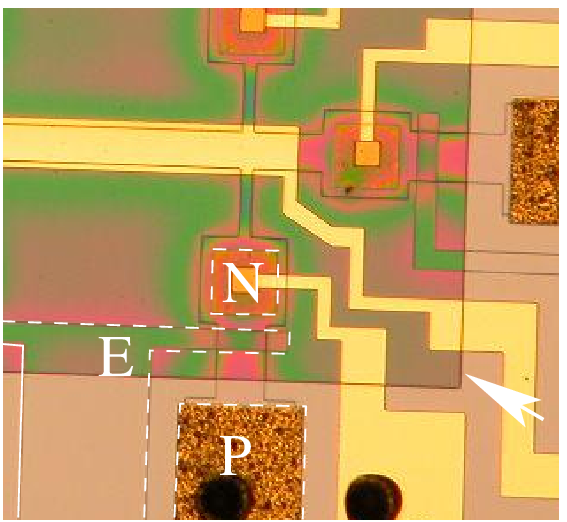}
\end{center}
\caption{\label{devicephoto}
(Color online) Photograph of a device showing the layout of the  n-contacts (N), p-contacts (P)
and isolation etch (E).
Openings in the polyimide(arrowhead)
allow bond pad metallisation to reach n-contacts.
}
\end{figure}

Diffused Au-Be alloy is used
to contact the hole layer and a carefully controlled  isolation etch is introduced between each pair of n and p contacts
(Figs\ref{deviceschematic}\&\ref{devicephoto}). The etch removes sufficient GaAs to depopulate
the inverted electron QW but leaves the lower (hole) QW unaffected. Fully independent
contacts are thus achieved without the need of any depletion gates, Focussed Ion Beam techniques or
shadow masking during MBE growth.

The devices were grown by MBE on  $<$100$>$ SI GaAs substrates.
Degradation of the hole gas mobility associated with diffusion of the Be dopant towards the `inverted' interface was reduced by
lowering the substrate temperature during the growth of the Be doped layer. A temperature of around 475$^\circ$C was found to be
optimal, giving  an improvement in mobility of one order of magnitude over wafers grown at 550$^\circ$C. The details
of the observed variation of 2DHG mobility with growth temperature will be reported elsewhere.\\
The devices are patterned into standard Hall-bar geometries ( 60$\mu$m wide with an aspect
ratio of 1:8.3, see figure \ref{devicephoto}), with six independent contacts to each layer.
In regions away from the contact (the Hall bar) the InAs
is completely removed with a selective etchant. ({\it e.g.} conc. HCl or  Succinic Acid in Ammonia).
 A sufficient amount of the doped GaAs cap is also removed such that the 2DEG is confined only
to a layer in an approximately
triangular well over the regions of the device away from the contacts. The bandstructure of
Fig.\ref{deviceschematic} depicts this situation.
In the simplest mode of operation, reported here, a single voltage bias between any one
pair of n and p type contacts, is used to induce the electrons. A threshold voltage
slightly higher than the bandgap (1.52V) of GaAs is required for the onset
of accumulation of electrons.
Both type of  contacts and carrier densities are stable and the devices show
reproducible behaviour
over several cooldowns from room temp to 300mK.
A polyimide layer is used to protect the sidewalls of the mesa, such that the
Ti/Au metallisation used to contact the InAs ohmics do not leak into the hole-layer.
Bonding directly to n-type contacts is avoided to prevent possible damage to the structure.
Using this procedure we have successfully produced
several devices with leakage currents $<$100pA at  interlayer bias $>$1.6V.
The contacting  scheme does not depend on the
precise depth of the 2DEG/2DHG and would remain effective if the barrier separating the
two layers is reduced further.

\begin{figure}[h]
\begin{center}
\includegraphics[width=8.5cm]{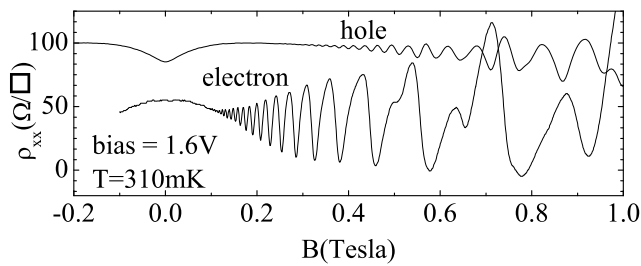}
\end{center}
\caption{\label{SdHoscillations}
Shubnikov de-Haas oscillations obtained using the independent ohmic contacts,
at interlayer bias V$_{eh}$=1.600V. The holes show a positive low-field magnetoresistance.}
\end{figure}

\begin{figure}[h]
\begin{center}
\includegraphics[width=8.5cm]{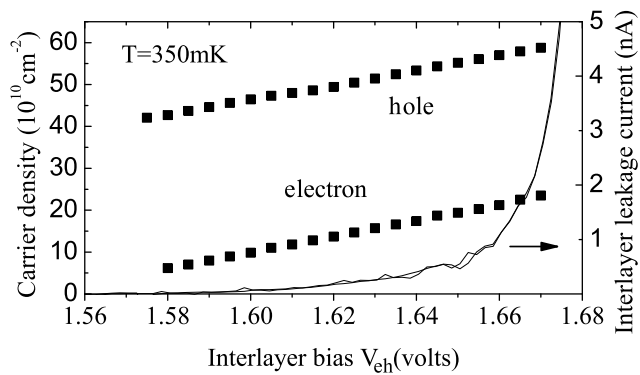}
\end{center}
\caption{\label{40nmNPvsveh}Variation of carrier densities with interlayer bias.}
\end{figure}

Typical Shubnikov de-Haas oscillations obtained from a device with a 15nm hole QW and 25nm
barrier are shown in Fig \ref{SdHoscillations}.
% The electron and hole carrier densities can be measured from
% either the Hall voltage or the period of SdH oscillations.
The quantum scattering times of the electrons measured
from a dingle plot of the oscillation amplitudes with magnetic field (not shown) give
$\tau_{q}$=1-2ps, comparable to values obtained from high mobility HEMTs grown in
the same MBE chamber. The carrier density and interlayer leakage measured at several
bias voltages are summarised in Fig \ref{40nmNPvsveh}.
 As the bias is increased, the  rate of increase in carrier density,  $dN/dV_{eh}$ is the same
for both the electrons and the holes, within experimental errors. The mean distance (d) between the electron and hole layers can
then be inferred from this capacitance ($\kappa\epsilon_{0}/d = edN/dV_{eh}$). The observed dependence of electron
density on the interlayer bias is N$_{e}$=1.91$\times$10$^{12}$(V$_{eh}$-V$_{0}$), where
V$_{0}$ is the threshold voltage and the densities are in cm$^{-2}$. This
gives a value of d=37nm, which is
in agreement with the peak to peak separation of the wavefunctions obtained from the
self-consistent calculations. The contact resistance of the InAs contacts was estimated
from the difference between the 2-probe and 4-probe resistances.
R$_{contact}\approx$5k$\Omega$ at an electron density of 5$\times$10$^{10}$cm$^{-2}$ and
falls to less than 400$\Omega$/contact at 2$\times$10$^{11}$cm$^{-2}$.
In the devices reported in this letter, the electron and hole densities cannot be
independently controlled. As shown in Fig \ref{40nmNPvsveh} the hole density is always higher than
the electron density, because of the contribution from the Be-dopants.
It is possible to incorporate a back-gate  to deplete the excess
hole density. This will be implemented in future devices.

Measurement of $\lq\lq$Coulomb drag" between a 2DEG and 2DHG in
closely spaced bilayers is of considerable current interest. We demonstrate that
our device can be successfully used to make these measurements.
%down to very low temperatures.
A low frequency
constant current is passed through one layer (the drive layer). Due to Coulomb
scattering between  carriers in two different layers, some momentum is transferred to
the other (drag) layer. Under open circuit conditions, this leads to a voltage appearing across
the drag layer. The magnitude of the voltage is a direct measure of the  interlayer
scattering rate.\cite{pogrebinski,price} Calculations based on linearised Boltzmann transport
equation give\cite{jauho}\\
\begin{eqnarray}
\nonumber
\rho_{drag}=\frac{\hbar}{e^{2}}\frac{2\pi{e^{2}}}{4k_{B}Tn_{1}n_{2}}\int\frac{d\vec{k_{1}}}{(2\pi)^{2}}
\frac{d\vec{k_{2}}}{(2\pi)^{2}}\frac{d\vec{k_{1{'}}}}{(2\pi)^{2}}\times\\
|\phi(q)|^{2}q^{2}f_{1}f_{2}(1-f_{1{'}})(1-f_{2{'}})
\delta(\epsilon_1+\epsilon_2-\epsilon_{1^{'}}-\epsilon_{2^{'}})
\end{eqnarray}
where $\phi(q)$ is the Fourier transform of the screened Coulomb potential, the numbers denote the
layer indices, the primes refer to the final states and $f_{1,2}$ are the relevant Fermi functions.
As long as the system is in the linear response regime (i.e. drive currents are sufficiently
small) Onsager's reciprocity theorem requires that the resistance measured by interchanging
the current and voltage terminals should be unchanged. \cite{casimir}

\begin{figure}[h]
\begin{center}
\includegraphics[width=8.5cm]{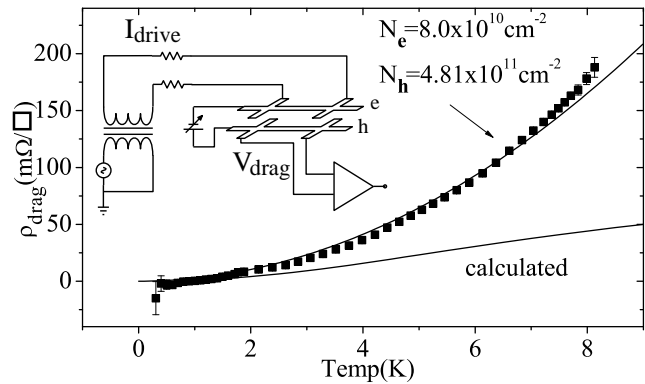}
\end{center}
\caption{\label{40nmdragplot}Coulomb drag in a 2DEG-2DHG structure with a 25nm barrier,
  measured using $\sim$ 100nA drive current at 7Hz. The line through data points
 is the best-fit to a T$^{2}$ dependence.}
\end{figure}

 In the data shown
(Fig \ref{40nmdragplot}), the error
bars represent the difference obtained by interchanging the roles of the drive and drag
layers. The differences are sufficiently small and $\rho_{drag}$
increases approximately as T$^{2}$. The power law can be qualitatively explained by
considering the phase-space available to the initial and final states in the
scattering process. The calculated magnitude of the drag resistance depends on the precise
form of the interaction used. Comparison with experimentally obtained
values of drag resistivity is an effective testing ground for theoretical  calculations of
the screened Coulomb potential.
Calculations based on the  Thomas-Fermi approximation
in a bilayer 2D system\cite{jauho,ando} overestimate the screening\cite{swierkowski}
and leads to the curve shown, for comparison.
This will be analysed in greater detail
elsewhere.

In conclusion, we have demonstrated a novel way of fabricating an electron-hole bilayer and
making independent contacts to each layer. The shallow n-type contacts do not require any
annealing and the samples can be patterned into Hall-bars. Standard MBE growth and
photolithographic techniques are used. Four-terminal measurements at 300mK as well as
Coulomb drag experiments have been made successfully  using these contacts.

{\bf Acknowledgements} This work was funded by EPSRC, UK.

\end{document}